\documentclass[conference]{IEEEtran}

\usepackage[letterpaper, left=1in, right=1in, bottom=1in, top=0.75in]{geometry}

\usepackage[pdftex]{graphicx}
\graphicspath{{images/}}
\usepackage[absolute,showboxes]{textpos}

\usepackage[cmex10]{amsmath}
\usepackage{mathtools}
\usepackage{amsfonts}
\usepackage{amssymb}
\usepackage{cite}

\usepackage{caption}
\usepackage{subcaption}

\usepackage{color}

\DeclareMathOperator{\Prob}{\mathsf{P}}
\DeclareMathOperator{\E}{\mathsf{E}}

\usepackage[absolute,showboxes]{textpos}

\TPMargin{5pt}

\usepackage{flushend}

\IEEEoverridecommandlockouts

\begin{document}
	
\title{Enhancing Cellular M2M Random Access with Binary Countdown Contention Resolution\thanks{This work has been funded in part by the German Research Foundation (DFG) grant KE1863/5-1 as part of the SPP 1914 CPN.}\vspace{-0.2cm}}

\author{\IEEEauthorblockN{Mikhail Vilgelm, Sergio Rueda Li\~nares, Wolfgang Kellerer}
\IEEEauthorblockA{Technical University of Munich, Germany\\
Email: \{mikhail.vilgelm, sergio.rueda, wolfgang.kellerer\}@tum.de}\\\vspace{-1cm}}

\maketitle
\begin{abstract}
Accommodating Machine-to-Machine applications and their requirements is one of the challenges on the way from LTE towards 5G networks. The envisioned high density of devices, alongside with their sporadic and synchronized transmission patterns, might create signaling storms and overload in the current LTE network. Here, we address the notorious random access (RA) challenge, namely, scalability of the radio link connection establishment protocol in LTE networks. We revisit the binary countdown technique for contention resolution (BCCR), and apply it to the LTE RA procedure. We analytically investigate the performance gains and trade-offs of applying BCCR in LTE. We further simulatively compare BCCR RA with the state-of-the-art RA techniques, and demonstrate its advantages in terms of delay and throughput.
\end{abstract}
\IEEEpeerreviewmaketitle

\section{Introduction}
\label{sec:intro}
With the saturating smartphone market, telecommunication vendors and operators are looking into new sources of revenue~\cite{osseiran2014scenarios_short}. One of these sources is a growing market of Machine-to-Machine (M2M) communications, devices operating without or with minimum human interaction. M2M devices are envisioned to create multiple design challenges, when fully integrated in the state-of-the-art cellular networks.
Key challenges are supporting high device density and diverse latency and reliability requirements of M2M devices~\cite{osseiran2014scenarios_short}. The diversity challenge, on the one hand, leads to the challenge of supporting quality of service (QoS) and differentiation between different M2M services. On the other hand, supporting high density of devices requires scalable network and radio layer protocols. For the current state-of-the-art systems, such as Long Term Evolution (LTE), it has been shown in multiple studies that the scalability can become an issue especially for the signaling procedures. For example, it has been shown that uneven growth of data and signaling might cause signaling storms, and high connection establishment delay~\cite{nokiaSignaling,osseiran2014scenarios_short,TR37868}.

In particular, the previous work has shown that LTE network is prone to an overload during random access (RA) procedure, if a large number of UEs with infrequent small data transmissions is present~\cite{TR37868,Mikhail2017Access}. Because of the high overhead of maintaining radio bearer, such UEs  must re-connect to the network prior to almost every transmission, hence, constantly loading Random Access Channel (RACH). The problem becomes even more amplified in the case of semi-synchronous, ``bursty'' arrivals of RA requests. This behavior is typical for many M2M applications~\cite{TR37868}, e.g., if multiple redundant sensors are reacting to the same emergency event. In addition to overload-induced throughput degradation, contention-based protocol nature of RA procedure makes explicit prioritization, necessary to distinguish different QoS classes, a challenging task.
\begin{figure}
	\centering
	\includegraphics[width=0.75\linewidth]{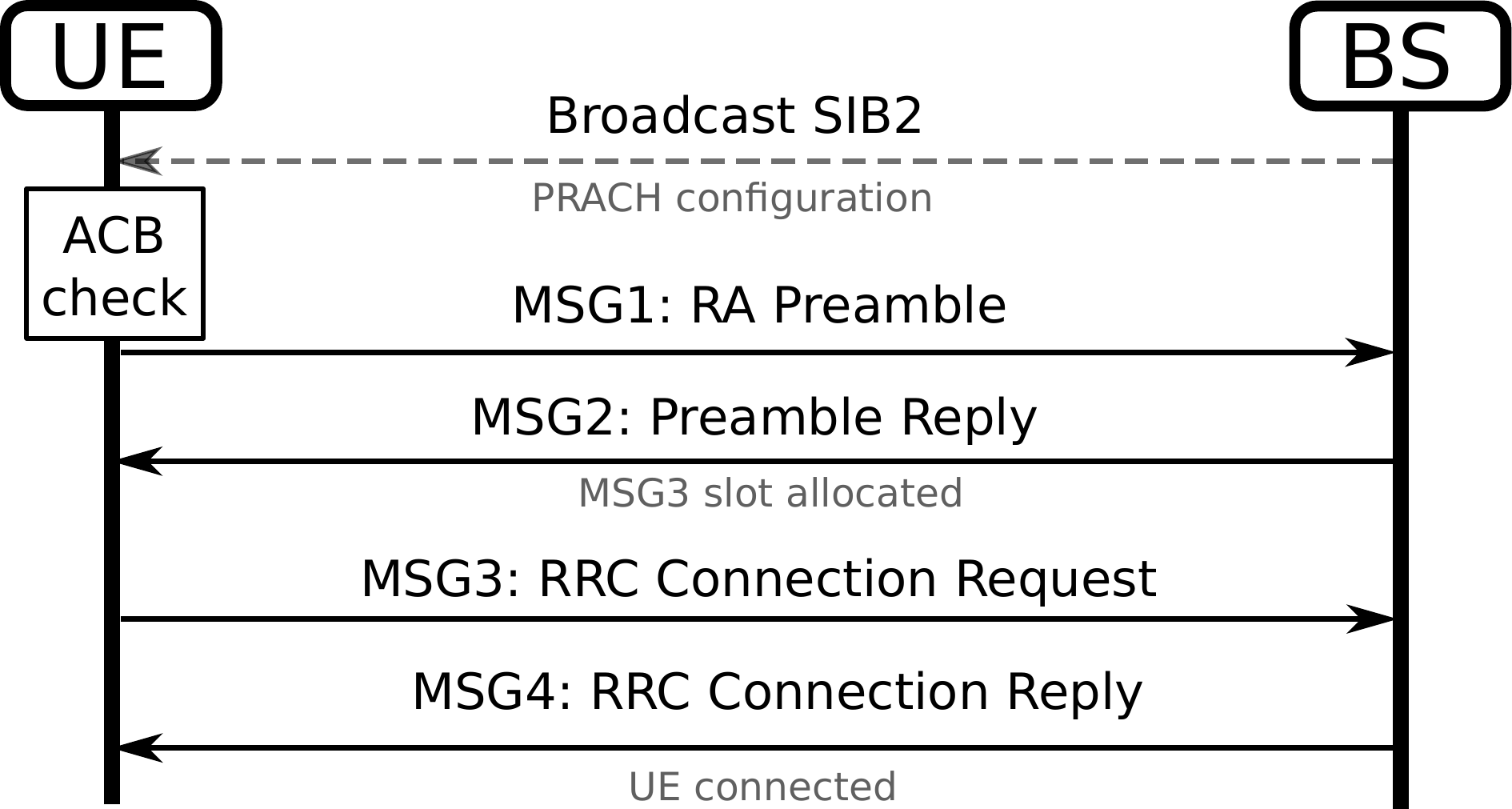}
	\caption{Standard protocol for LTE Random Access Procedure.}
	\label{fig:msg3bpr}
	\vspace{-0.7cm}
\end{figure}

The RA challenge has attracted significant research attention in the last years, and a multitude of solutions for increasing throughput and providing prioritization have been introduced~\cite{Mikhail2017Access,7870628,laya2014random,duand,madueno2015massive}, which we briefly review in Sec.~\ref{sec:relatedwork}.
However, practically all of them focus on the first part of the procedure, namely the preamble contention. In contrast to them, in our approach, we \textit{accept preamble collisions}, and focus on the contention resolution prior to connection request, third step of the RA procedure (see Fig.~\ref{fig:msg3bpr}). This makes our approach orthogonal to most of the state-of-the-art, so that it is possible to combine our protocol with any preamble contention optimization technique. Also, our approach is applicable both for improving the throughput, and providing prioritization during RA procedure.

In this paper, we revisit binary countdown contention resolution (BCCR) protocol, which has been a common collision preventing technique in bus and powerline communications~\cite{gehrsitz2014priority}, dating back to the works of Mok~\textit{et. al}~\cite{mok1979distributed}. We apply BCCR to cellular random access, and show how it can be used to improve RACH throughput, and provide prioritization. We analytically model BCCR performance, and evaluate the overhead and trade-offs of applying it in LTE. Simulatively, we compare BCCR to state-of-the-art RA in terms of delay and throughput.

The remainder of the paper is structured as follows. The background of LTE RACH is reviewed in Sec.~\ref{sec:background}. We introduce BCCR in Sec.~\ref{sec:brp}, and study its performance analytically in Sec.~\ref{sec:analysis} and simulatively in Sec.~\ref{sec:evaluation}. The paper is concluded with Sec.~\ref{sec:conclusions}.

\section{Background and Related Work}
\label{sec:background}

\vspace{-0.1cm}
\subsection{Random Access Procedure in LTE}

The conventional four-step Random Access Procedure is depicted in Fig.~\ref{fig:msg3bpr}. It starts with a user equipment (UE) retrieving Physical Random Access Channel configuration from base station (BS) system information broadcast (SIB2). The configuration typically contains time-frequency location of the PRACH slot, contention parameters, such as barring probability, and the set of available preambles $\mathcal{M}$ with $M=|\mathcal{M}|$. Typically, there are $M\leq54$ preambles available for contention-based RA procedure. As a part of the access class barring (ACB) procedure, the barring probability $\Prob_b$ is used by every UE to decide if it starts the RA procedure in the upcoming PRACH slot. With probability $\Prob_b$, UE  postpones the access until the next slot, i.e., it is considered barred.

Any non-barred UE proceeds with sending a randomly chosen preamble from the set $\mathcal{M}$ as MSG1 in the PRACH slot. After RA preamble reception, BS replies with preamble response (MSG2), containing the timing and location of the sub-frame for RRC Connection Request (MSG3) for \textit{every} received (``activated'') preamble. Since a preamble does not contain any UE identification information, any preamble chosen by at least one UE is considered activated. Hence, if two or more UEs choose the same preamble as MSG1, they are allocated the same MSG3 slot, which leads to collisions. If Connection Requests collide, no MSG4 is received from the BS, and the collided UEs re-attempt sending the preambles after a random back-off time. If, however, a UE has chosen unique preamble, its MSG3 is successfully received by BS, leading to the successful RA procedure completion with RRC Connection Reply (MSG4) from BS. 

\vspace{-0.1cm}
\subsection{Related Work}
\label{sec:relatedwork}

In this section, we briefly review state-of-the-art on RACH. For the in-depth overview, we refer the reader to~\cite{Mikhail2017Access,7870628,laya2014random}. First challenge of RA procedure is that its collision allowing protocol is prone to performance degradation under high load, especially in the case of semi-synchronous M2M arrivals~\cite{TR37868}. RA procedure is typically modeled as a multi-channel slotted ALOHA (where preambles are channels)~\cite{wei2015modeling,duand}. 
Multiple solutions have been proposed to boost the scalability of RACH, often based on the earlier approaches to single channel ALOHA throughput enhancements. For instance, dynamic access barring~\cite{duand} for smoothing the burst load, or Q-ary tree splitting algorithms for fast collision resolution~\cite{madueno2015massive,7870628}. Additionally, back-off and duty cycle adjustments have been explored~\cite{laya2014random}.
Some solutions propose a more invasive re-design of the random access channel, e.g., with distributed queuing~\cite{laya2014random}. 

Besides improving the scalability, the other RA challenge is to provide efficient prioritization. Typical prioritization techniques with contention parameters do not provide sufficient isolation between the priority classes, while resource-separation based techniques rely on the accurate load estimation, which might not be available. In the contrast to state-of-the-art, proposed here protocol does not include manipulating preamble contention. Instead, we attempt to resolve MSG3 after the preamble collision, applying binary countdown technique~\cite{mok1979distributed}.

\section{Proposal: Binary Countdown Contention Resolution (BCCR) Protocol}
\label{sec:brp}
\begin{figure*}[]
	\centering
	\includegraphics[width=\textwidth]{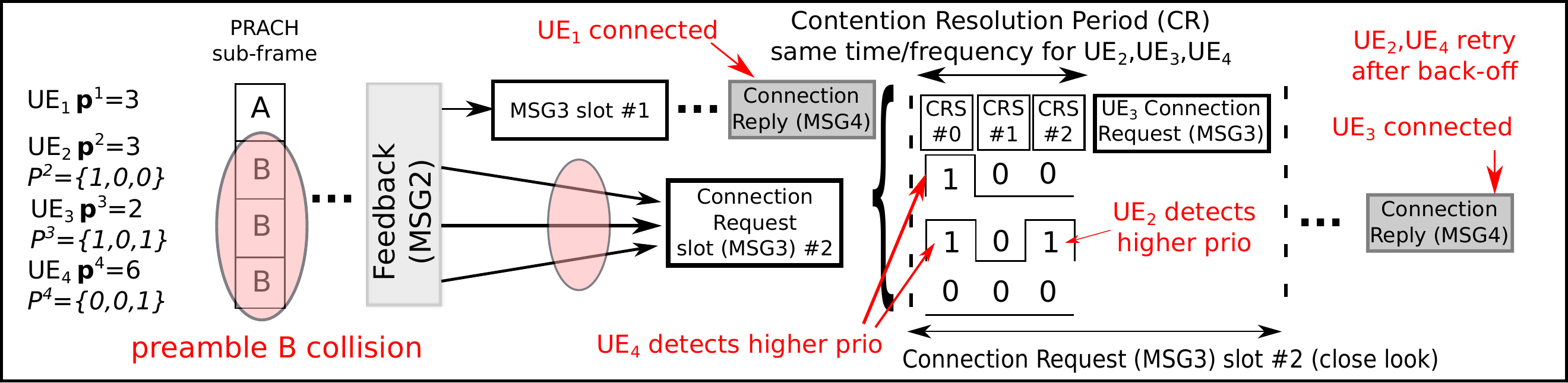}
	\caption{Exemplary operation of RA procedure with Binary Countdown Contention Resolution.}
	\label{fig:procedure}
	\vspace{-0.7cm}
\end{figure*}

The proposed approach relies on a binary countdown protocol for resolving a contention and preventing collisions~\cite{mok1979distributed}.
In the LTE case, we aim to prevent a MSG3 collision, after a preamble collision has occurred.

For the protocol operation, we need a modified MSG3 slot structure, see Fig.~\ref{fig:procedure}, where the actual MSG3 sending is preceded with a contention resolution (CR) period, consisting of $k$ CR micro-slots (CRSs), with $l=2^k$ priority levels. Prior to contention, every UE $i$ locally chooses a \textit{priority} $\mathbf{p}^i$, represented as a binary sequence $P^i=[p^i_0,\dots,p^i_j,\dots,p^i_{k-1}]$, with $p^i_j\in\{0,1\}$, $0\leq j\leq k-1$, such that $P^i$ is a value of $(l-1-\mathbf{p}^i)$ in base-2 numeral system. For example, for $k=2$ micro-slots, highest priority $\mathbf{p}^i=0$ is mapped onto sequence $P^i=[1,1]$, and lowest priority $\mathbf{p}^i=3$ is represented as $P^i=[0,0]$. We adhere to conventional notation that highest possible priority is always 0. We will discuss the possible options on how to choose priorities in~\ref{sec:application}.

Then, the priority sequences are used to decide individual UE's behavior during the CR period: in a given $j$th CR slot, UE $i$ is either broadcasting signal if $p^i_j=1$, or listening to the medium (if $p^i_j=0$). If a UE detects a signal in a slot while listening (meaning, higher priority UE is present), it immediately stops the contention resolution, and goes directly for back-off. Winning UE proceeds with sending MSG3 to BS. The procedure ensures that (1) provided the highest priority UE is unique, it is guaranteed to send MSG3 without collision, and (2) other UEs ``loosing'' in the contention, start the back-off earlier, not waiting for MSG4 timeout.

\vspace{-0.1cm}
\subsection{BCCR Operation Example}
In this section, we demonstrate the operation of BCCR on a simple example, illustrated in Fig.~\ref{fig:procedure}. Given are $k=3$ CRS with up to $l=2^k=8$ priorities, and four $\text{UE}_i$, $i\in\{1,2,3,4\}$ with respective local priorities $\mathbf{p}^1=\mathbf{p}^2=3$, $\mathbf{p}^3=2$, $\mathbf{p}^4=6$. Hence, their binary sequences are $P^1=P^2=[1,0,0]$, $P^3=[1,0,1]$, $P^4=[0,0,1]$. 

Assume that all four UEs passed ACB check, and are eligible to compete in a given PRACH slot. $\text{UE}_1$ chooses preamble A, and $\text{UE}_2$, $\text{UE}_3$, $\text{UE}_4$ choose preamble B. BS detects two activated preambles, and respectively allocates two MSG3 slots: \#1 for preamble A ($\text{UE}_1$), and \#2 for preamble B ($\text{UE}_2$,$\text{UE}_3$,$\text{UE}_4$). Since preamble A is unique, $\text{UE}_1$ sends MSG3 without collision, and goes into connected states after the MSG4 feedback from BS. The remaining UEs are attempting to resolve the contention using BCCR, before sending their MSG3's. In the $0$th CRS, only $\text{UE}_2$ and $\text{UE}_3$ are transmitting signal, while $\text{UE}_4$ is listening ($p^2_0=p^3_0=1$ and $p^4_0=0$). Since $\text{UE}_4$ is detecting a non-empty signal in the first CRS, it concludes that higher priority UE is present, and, hence, drops the attempt and goes for back-off. Importantly, this implies that $\text{UE}_4$ \textit{does not broadcast} in the $2$nd CRS, although $p^4_2=1$.
At the $1$st CRS, no UE broadcasts ($p^2_0=p^3_0=0$). Finally, during the $2$nd CRS, only $\text{UE}_3$ broadcasts the signal, while $\text{UE}_2$ is listening ($p^2_0=0$, $p^3_0=1$). Since the signal from $\text{UE}_3$ is present, $\text{UE}_2$ does not proceed with transmission and backs off. As a result, $\text{UE}_3$ wins the contention and continues with Connection Request transmission. 

To summarize, BCCR has increased the throughput of RA procedure from 1 UE to 2 UEs per PRACH slot, at the expense to adding overhead of $k=3$ CRS per every MSG3 slot. Note that, since BS cannot distinguish between collided preamble and non-collided preamble, CR period has to precede every MSG3, even if no preamble collision was present. We will return to this trade-off during the overhead analysis in~\ref{sec:tradeoff}.

\vspace{-0.1cm}
\subsection{BCCR Priority Assignment}
\label{sec:application}

Conventionally, binary countdown sequences and respective priorities are assigned to the users  based on their application type. This is typical for binary countdown in CAN bus, because of the inherently hierarchical functioning of the system; i.e., nodes can be easily distinguished by the priority of their function~\cite{gehrsitz2014priority}. In a similar way, LTE UE's priority can be assigned based on Quality of Service class, on a per-user or even per-flow basis. On top of prioritization, full contention-free access could be potentially achieved, if a sequence is prepended with a unique user identifier. However, this might be hard to implement in LTE in practice, since it requires many BCCR slots, and raises fairness issues.

Finally, the main priority assignment scenario we consider in this paper is \textit{uniformly random choice of priorities}. This gives another ``channel'' dimension for multichannel ALOHA, similarly to preambles, which allows to improve the overall throughput of the LTE RA. Randomization and prioritization can be even implemented together, at the expense of longer CR period.

\vspace{-0.1cm}
\subsection{Implementation in LTE}

In order to ensure a seamless integration in the LTE system functioning, we propose that BCCR takes place in CRS of the same duration as one LTE symbol, $t_{\text{CRS}}=66.67 \mu$s. In that way, cyclic prefix prior to each symbol period is maintained as guard period for robustness.

In LTE, synchronization of UEs and BS is handled by a specific Timing Advance (TA) for each device, compensating both the heterogeneity in the uplink propagation delay and its time variability.
However, for RA, if multiple UEs are contending for MSG3, they all receive the same TA instructions via MSG2, and thus BCCR in the LTE RA procedure is intrinsically unsynchronized. Therefore, our proposed approach is not to transmit during the entire contention resolution slot time duration, $t_{\text{CRS}}$, but rather only during its first part, with duration $t^\prime_{\text{CRS}}$. For the pair of devices $\text{UE}_i, \text{UE}_j$, $i\neq j$, contending to send MSG3 over the same PUSCH resources, we denote $d_i, d_j$ as their respective distances to the BS and $d_{i,j}$ as the distance between each other. For the sake of robustness, it is important to ensure that every UE is able to hear the broadcast from all other contending UEs, arriving entirely within $t_{\text{CRS}}$. Thus, the worst case scenario is when the first contending UE starts transmitting (closest to BS), and has to wait to hear the last UE (furthest). Denoting closest UE as $i=1$ and furthest as $j=2$, this restriction is expressed as:
\begin{equation} \label{eq:cond1}
t_{\text{CRS}} \geq (d_2 - d_1)/c + t^\prime_{\text{CRS}} + d_{1,2}/c
\end{equation}
being $c$ the signal propagation speed, approximately equal to the speed of light. Furthermore, given the triangle inequality $d_j - d_i \leq d_{i,j}$, we can obtain a more restrictive but simpler condition to work with, satisfying Eqn.~\eqref{eq:cond1}: $t_{\text{CRS}}  \geq t^\prime_{\text{CRS}} + 2d_{1,2}/c$.

This allows us to calculate the minimum BCCR hearing diameter as a function of the ratio $t^\prime_{\text{CRS}}/t_{\text{CRS}}$. Note that, assuming a fixed transmission power, the higher the ratio $ t^\prime_{\text{CRS}}/t_{\text{CRS}}$ is, the greater the robustness against SINR degradation. E.g., for a conservative ratio of 0.9, we obtain a hearing distance of approximately 1 km; i.e., every device is able to contend at least with every other device less than 1 km away. Furthermore, it is important to note that aforementioned "bursty" arrivals are likely to be geographically as well as temporally correlated. Thus, moderate values of the hearing distance are likely to suffice in such scenarios.

\vspace{-0.1cm}
\section{Analysis}
\label{sec:analysis}

In this section, we analyze contention resolution capability of randomized priority assignment in BCCR, and evaluate the overhead impact on effective throughput.

We consider a single PRACH slot, with $n$ UEs successfully passing access barring, and contending for $M$ preambles. We further denote the number of available priority levels as $l$, and the number of CR slots as $k$. In the previous work, it has been shown that, given $n$ and $M$, the expected number of successfully connected UEs (without considering BCCR) $\Delta n$, is computed as~\cite{wei2015modeling}:
\begin{equation}
\Delta n = n\left(1-\frac{1}{M}\right)^{n-1}.
\label{eqn:deltan}
\end{equation}

To quantify the impact of $l$ priority levels, we need to compute (1) how many of $M$ preambles have been involved in a collision $M_C$, and (2) what is the expected collision size for a single preamble $s^\prime$.

To compute $\E[M_C]$, we note that $\E[M_C] = M - \Delta n - \E[M_I]$, where $M_I$ is the number of idle preambles, i.e., preambles not chosen by any UE. Probability of $M_{I}=m$ out of $M$ preambles to be idle is:
\begin{align}
\Prob[M_{I}=&m] = \binom{M}{m}\left(\frac{M-m}{M}\right)^n,\text{ hence,}\nonumber\\
\E[M_I] =& \sum_{m=0}^{M}m\Prob[M_{I}=m]=\nonumber\\
=& \sum_{m=0}^{M}m\binom{M}{m}\left(\frac{M-m}{M}\right)^n
= M\left(1-\frac{1}{M}\right)^{n}.\nonumber
\end{align}

\begin{figure}[t!]
	\centering
	\begin{subfigure}[t]{\linewidth}
		\centering
		\includegraphics[width=\linewidth, trim=0.0 0.4cm 0.0 0.5cm, clip=true]{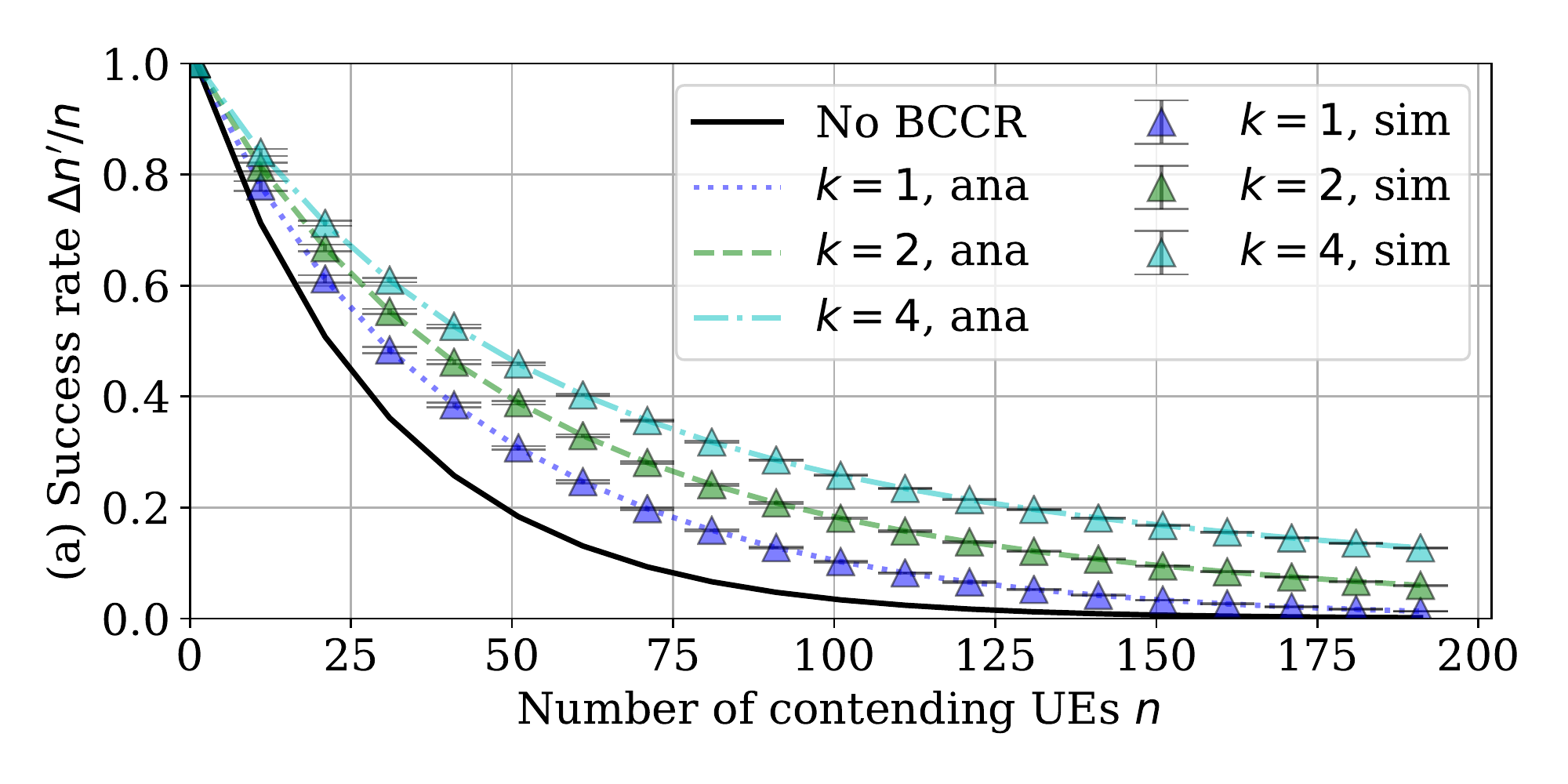}
		\vspace{-0.3cm}
	\end{subfigure}%
	
	\begin{subfigure}[t]{\linewidth}
		\centering
		\includegraphics[width=\linewidth, trim=0.0 0.4cm 0.0 0.5cm, clip=true]{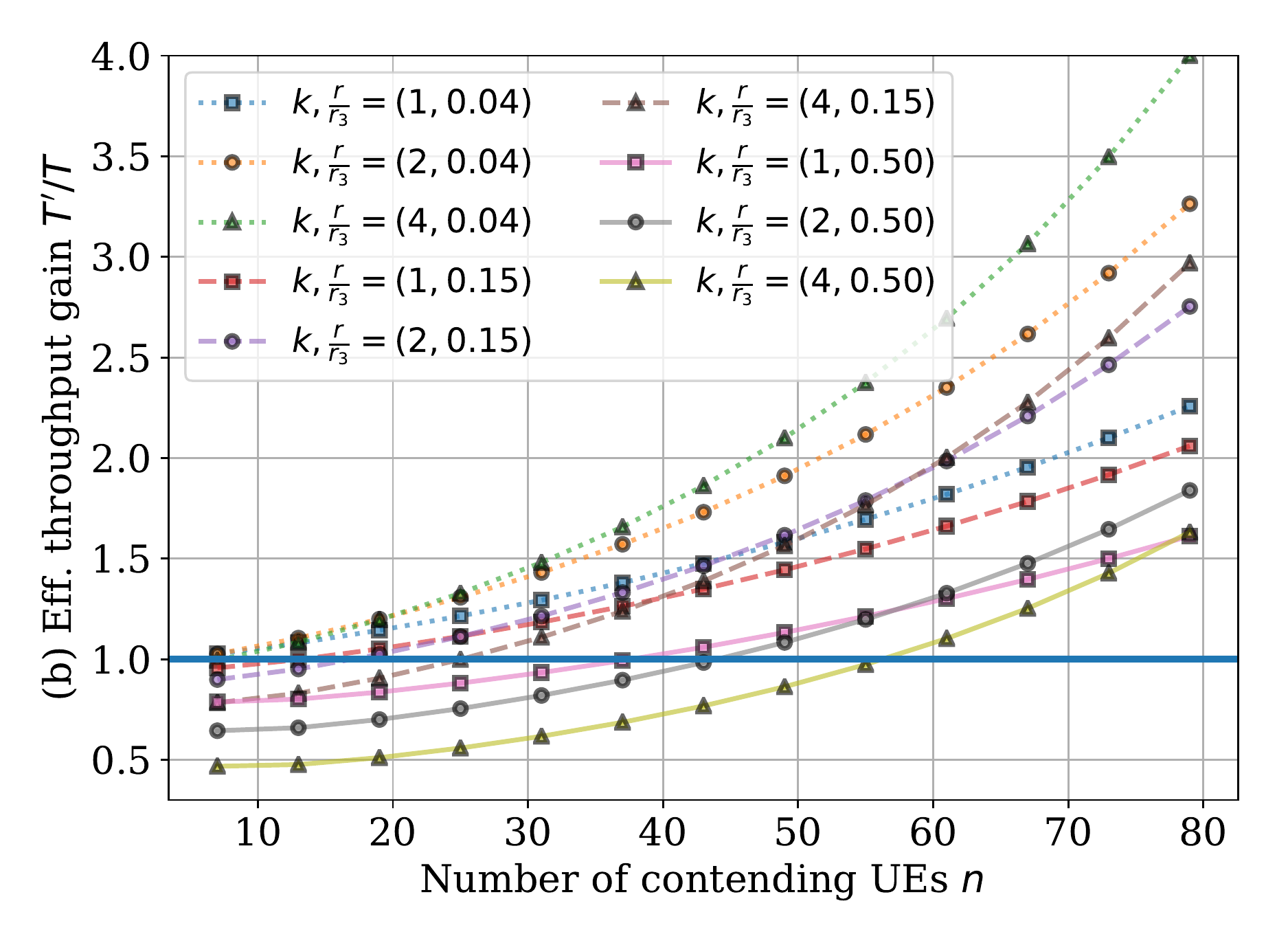}
		\vspace{-1cm}
	\end{subfigure}
	\caption{Above: (a) UE success ratio $\Delta n^\prime/n$; Below: (b) Effective throughput gain $T^\prime/T$ vs. number of contending (non-barred) UEs in a PRACH slot $n$. $k\in\{1,2,4\}$ CRSs; $M=30$; 95~\% confidence intervals.}
	\label{fig:successratio}
	\vspace{-0.6cm}
\end{figure}

\begin{figure*}[t!]
	\centering
	\includegraphics[width=\linewidth, trim=0.0 0.62cm 0.0 0.6cm, clip=true]{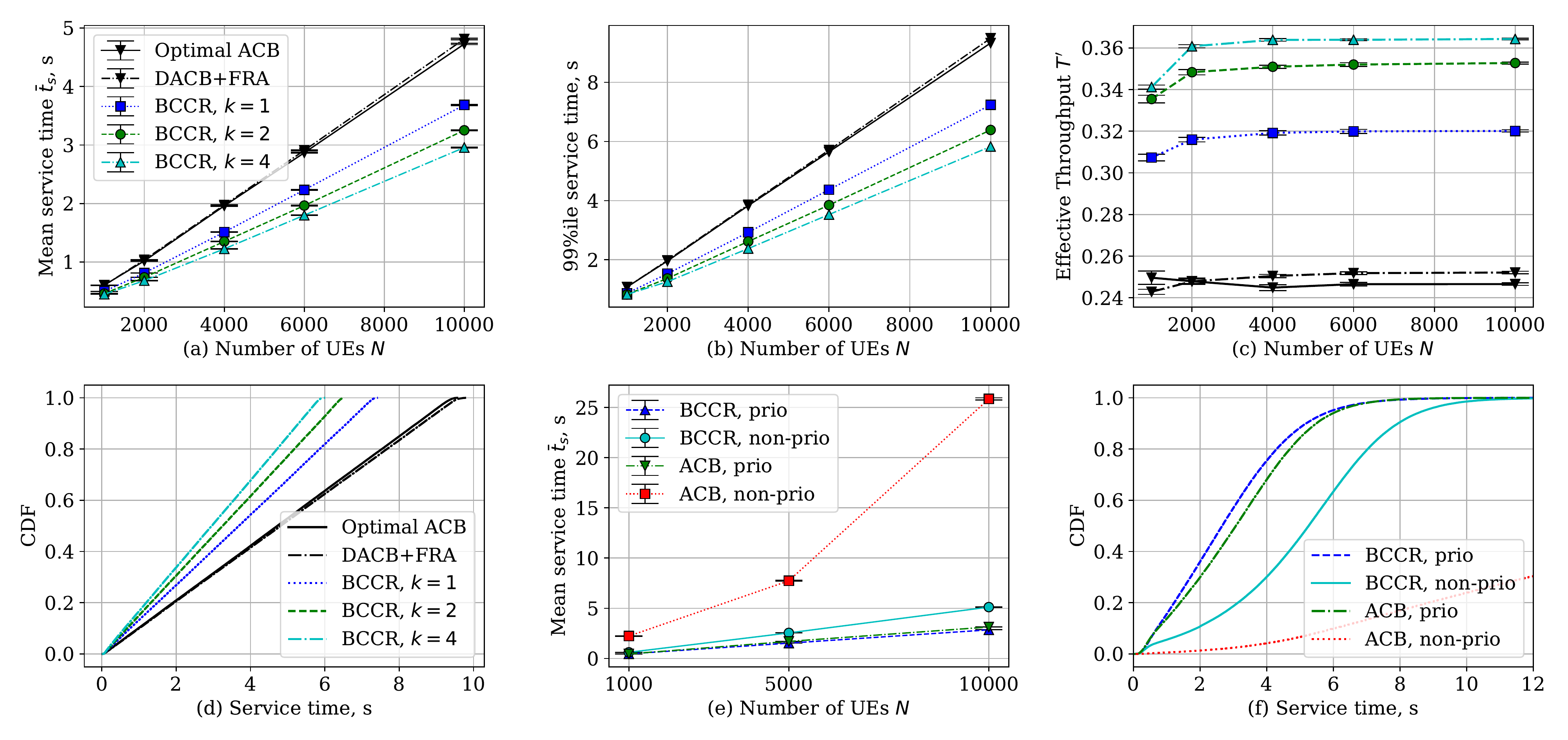}
	\caption{Randomized priorities: (a) Mean and (b) 99\%ile of the service time; (c) effective throughput $T$ vs. number of UEs $N$; (d) Service time CDF for $N=10000$. Two-class prioritization: (e) mean service time $\bar{t}_s$ vs. number of UEs $N$; (f) Service time CDF for $N=10000$; $T_A=1$~s, $\alpha=3$, $\beta=4$; $k\in\{1,2,4\}$ CRSs; D-ACB with FRA~\cite{duand}; $M=30$; $95$~\% conf. int.}
	\label{fig:overallperformance}
	\vspace{-0.6cm} 
\end{figure*}

Expected number of preambles involved in the collision is, thus, $M-\Delta n - \E[M_I]$, and expected number of collided UEs is $n-\Delta n$. Hence, we obtain an expected collision size as:
\begin{eqnarray}
s^\prime = \frac{n-\Delta n}{M-\Delta n - \E[M_I]}
= \frac{n-\Delta n}{M-\Delta n-M\left(1-\frac{1}{M}\right)^{n}}.\nonumber
\end{eqnarray}

Given an expected preamble collision size $s^\prime$, number of BCCR priorities $l$, the successful outcome of contention resolution on MSG3 occurs if any UE chooses a priority higher than all other UE. Since any priority $p\in\{0,\dots,l-2\}$ can be the highest, total probability can be expressed as:
\begin{eqnarray}
\Prob_{\text{CR}} = \Prob[\text{highest priority level is unique}|s^\prime] = \nonumber\\ = \sum_{p=0}^{l-2}\frac{s^\prime}{l}\left(1-\frac{p+1}{l}\right)^{s^\prime-1}.
\end{eqnarray}

Since any of the preamble collisions can be resolved via BCCR with the probability of $\Prob_{\text{CR}}$, total number of successful UEs in a given slot is given by:
\begin{eqnarray}
\Delta n^\prime
=& \Delta n + (M-\Delta n-\E[M_I])\Prob_{\text{CR}}=\label{eqn:deltaprime}
\end{eqnarray}
\vspace{-1.2cm}

\begin{eqnarray}
= n\left(1-\frac{1}{M}\right)^{n-1}+\left(M-\Delta n-M\left(1-\frac{1}{M}\right)^{n}\right)\times\nonumber\\
\times
\sum_{p=0}^{l-2}\frac{s^\prime}{l}\left(1-\frac{p+1}{l}\right)^{s^\prime-1}.\nonumber
\end{eqnarray}

We plot the results success ratio $\Delta n^\prime/n$ for $k\in\{1,2,4\}$ CRS, and compare it to the base line $\Delta n/n$ in Fig.~\ref{fig:successratio}(a) for $n=[2,...,195]$ UEs. We observe that the analysis allows accurate prediction of success ratio via Eqn.~\eqref{eqn:deltaprime}. Also, we observe that BCCR significantly increases success ratio, even in the case of only one CRS, e.g., from $0.19$ to $0.32$ for $n=50$.

\vspace{-0.1cm}
\subsection{Overhead and Effective Throughput}
\label{sec:tradeoff}
Introducing BCCR in LTE RA procedure is also introducing overhead due to the resources reserved for CRSs. BCCR operating with $l$ priority levels requires $k=\lceil\log_2l\rceil$ CRSs. Given the amount of resources (time$\times$frequency) required for single MSG3 as $r_3$, and amount of resources for all MSG1's as $R_1$, overall amount of uplink resources spent on a single RA procedure cycle is given as:
\begin{equation}
R_{\sum} = R_1 + r_3\underbrace{\left(M - M\left(1-1/M\right)^{n}\right)}_{\text{all non-idle preambles}}.
\end{equation}

Define the resources per contention resolution slot as $r$. Hence, for BCCR RA with $l$ priority levels:
\begin{eqnarray}
R^\prime_{\sum} = R_1 + \left(\lceil\log_2l\rceil r+r_3\right)\left(M - M\left(1-1/M\right)^{n}\right)\nonumber
\end{eqnarray}

Define as \textit{effective throughput} $T=\Delta n/R_{\sum}$ the amount of successful RA requests normalized by the total number of resources spent, during a single RACH slot. Accordingly, for BCCR RA we get $T^\prime=\Delta n^\prime/R_{\sum}^\prime$. To evaluate the effect of the BCCR overhead, we look at the ratio $T^\prime/T$, characterizing the effective throughput \textit{gain}. The gain is shown in Fig.~\ref{fig:successratio}(b). We fixed $R_1=6$~RBs, since PRACH typically occupies $6$~RBs in LTE. The value for $r_3$ might in general vary due to protocol implementation and channel variation. Here, we assume $r_3=2$~RBs~\cite{jang2016non}.

We show the gain for three different per slot overhead values $r/r_3$: $0.04$ (proposed option of $1$ symbol per BCCR slot), medium value $0.15$, and very high value $0.5$ (for $r_3=2$, it corresponds to 1 RB long BCCR slot). We observe from Fig.~\ref{fig:successratio}(b) that for the proposed BCCR implementation, effective throughput gain exists even for low number of contending UEs. However, for very high BCCR slot duration $r/r_3=0.5$, BCCR usage only makes sense if high number of contending UEs is present $n\geq 36$. The higher is the number of priority resolution slots, the higher grows the gain with the number of UEs $n$, however, also higher is the minimum number of UEs where gain is larger than $1$.

The analysis methodology proposed here is generalizable and Eqn.~\eqref{eqn:deltaprime} can be used to efficiently characterize both steady-state or transient state delay and throughput of the multi-channel ALOHA with BCCR, e.g., using the methodology proposed in~\cite{wei2015modeling}.
\section{Performance Evaluation}
\label{sec:evaluation}

In this section, we evaluate delay and throughput performance of BCCR RA for typical M2M burst arrival scenario~\cite{TR37868}. In~\ref{sec:randomized}, we study a single class RA and randomized binary sequence choice, and in~\ref{sec:prioritization} we compare prioritization using BCCR with prioritization using access class barring. 

\subsection{Evaluation Set-up}
We consider a single LTE cell with a BS and a number of UEs in its coverage. We study a scenario of an emergency event involving wake-up of $N$ UEs.

Such an emergency situation provokes involved UEs to re-connect to the BS over a short period of time $T_A$, referred to as \textit{activation time}. As proposed by 3GPP~\cite{TR37868}, such a traffic burst can be modeled by a beta distribution, where each UE starts the reconnection procedure at time $t$ with probability distribution $g(t) = \frac{t^{\alpha-1} (T_A-t)^{\beta-1} }{T_A^{\alpha+\beta-2} B (\alpha,\beta)}$, $0\leq t\leq T_A$, where $ B (\alpha,\beta)$ denotes the Beta function.

We study three performance metrics: mean service time $\bar{t}_s = \sum_{j=1}^{N}t_s^j / N$ (time until a UE successfully completes RA procedure) of all UEs, service time distribution, and effective RA throughput $T$ as defined in~\ref{sec:tradeoff}. Since we are targeting high criticality and reliability scenario, we additionally study 99\%ile of the service time, $P_{99}$.

Simulated model only captures MAC layer effects. Packets are assumed to be successfully received if no collision has occurred, furthermore, possible limitations of physical downlink shared channel are also not captured in the simulation. These effects might have a quantitative impact on the results, however, they would not bring a qualitative change, therefore, we omit them to make the evaluation more illustrative.

\subsection{Randomized Priority Sequences}
\label{sec:randomized}
In order to assess the performance improvement resulting from BCCR, a number of scenarios with varying number of CRS $k$ have been simulated and compared to the baseline of conventional RA. We study only the case of geometric back-off with access barring, applied in all cases regardless of the value of $k$. For comparison, we choose the dynamic access barring algorithm from~\cite{duand}. The algorithm relies on dynamically changing the barring factor, in order to keep the number of contending UEs around the optimal point of $M$ UEs per slot. We have implemented DACB with full state knowledge (less realistic scenario to provide an optimistic baseline), and DACB with state estimation (DACB+FRA). For BCCR evaluation, we also use DACB+FRA as a preamble contention optimization technique.

In Fig.~\ref{fig:overallperformance}(a-d) we observe how an increase in the number of priority levels notably enhances the system performance. We note that, for all $k$, both (a) the average service time and (b) the 99\%ile of the service time CDF display a linear increasing behaviour with respect to the number of UEs. In (c) we observe how the throughput $T$ saturates to a maximum value as the number of UEs increases, since the quasi-optimal barring factor prevents the system from overloading. Most interestingly, we see how an increased $k$ provokes a clear boost in the aforementioned maximum throughput of the system, especially with $k=4$ CRS. Fig.~\ref{fig:overallperformance}(d) shows the Service time CDF for the case $N = 10000$ UEs, where better throughput of the system with increasing $k$ is clearly made visible as an increasing slope in the CDFs.

\subsection{Prioritization via BCCR}
\label{sec:prioritization}

Here, we assess the possibility of using BCCR to prioritize certain traffic classes in high load condition. In Fig.~\ref{fig:overallperformance}(e,f), the results of prioritization of two classes (namely, ``prio'' and ``non-prio'') are compared for BCCR and ACB, with respect to the service time.
For BCCR, a common barring probability for both classes $\Prob_b^1 = \Prob_b^2$ was arbitrarily set. These barring probabilities were chosen to be increasing with the number of devices. To ensure fair comparison, we then set ACB parameters $\Prob_b^1$, $\Prob_b^2$ such that the 99\%ile for the prioritize class matches that of BCCR.

From Fig.~\ref{fig:overallperformance}(e) we observe that, to achieve similar ``prio'' class mean service time, ACB scheme significantly deteriorates the ``non-prio'' class performance, and BCCR results in up to $5$~times lower delay for ``non-prio'' class.  Fig.~\ref{fig:overallperformance}(f) illustrates the 99\%ile match between the CDFs of the ``prio'' class under both schemes. It also hints to the fact that one of the main reasons why the ACB throughput is so deteriorated compared to the BCCR scheme is the high barring factor that we need to apply to the ``non-prio'' class in order to reserve sufficient resources for the ``prio'' class. 

\section{Discussion and Conclusions}
\label{sec:conclusions}

In this paper, we have revisited binary countdown based contention resolution (BCCR), and applied it to LTE RACH. We have shown, that if BCCR is applied after the preamble collision, and prior to Connection Request (MSG3), it significantly reduces the amount of collisions, boosts the throughput and reduces the delay. Using the definition of effective throughput, we have introduced a framework for BCCR overhead evaluation, and demonstrated that the proposed BCCR is valuable and increases the effective throughput. Simulatively, we have compared RA delay of BCCR and the baseline dynamic access barring. Additionally, we have shown how BCCR can also be used for prioritization, and illustrated its superiority compared to access class barring.

As the effective throughput evaluation methodology in~\ref{sec:analysis} suggests, further work on BCCR RA includes developing a dynamic approach -- using the estimation of the number of collisions, number of CR slots can be adjusted to match the load. This will boost the performance of LTE RACH even further, at the same time reducing the overhead of static BCCR.

\bibliographystyle{IEEEtran}

\end{document}